\documentclass[aps,amsmath,amssymb,superscriptaddress,showpacs,twocolumn,floats,epsf]{revtex4}
\usepackage{amsfonts}
\usepackage{graphics}
\usepackage{graphicx}
\usepackage{hyperref}
\usepackage{color}

\begin{document}

\title{Diffusive Quantum Criticality in Three Dimensional Disordered Dirac Semimetals}

\author{Bitan Roy}
\affiliation{Condensed Matter Theory Center and Joint Quantum Institute, University of Maryland, College Park, Maryland 20742-4111, USA}

\author{S. Das Sarma}
\affiliation{Condensed Matter Theory Center and Joint Quantum Institute, University of Maryland, College Park, Maryland 20742-4111, USA}

\date{\today}

\begin{abstract}
Three dimensional Dirac semimetals are stable against weak potential disorder, but not against strong disorder. 
In the language of renormalization group, such stability stems from the irrelevance of weak disorder in the vicinity of the noninteracting Gaussian fixed point. However, beyond a threshold, potential disorder can take Dirac semimetals into a compressible diffusive metallic phase through a quantum phase transition (QPT), where density of states at zero energy, quasiparticle lifetime and metallic conductivity at $T=0$, are finite. Universal behavior of such unconventional QPT is described within the framework of an $\epsilon(=d-2)$-expansion near the lower critical dimension. Various exponents near this QCP are obtained after performing a two loop perturbative expansion in the vanishing replica limit and we demonstrate that the theory is renormalizable at least to two loop order.We argue that such QPT is always continuous in nature, share same university class with a similar transition driven by odd-parity disorder, critical exponents are independent of flavor number of Dirac femrions and thus can be germane to disordered Cd$_3$As$_2$, Na$_3$Bi. Scaling behaviors of various measurable quantities, such as specific heat, density of states across such QPT are proposed.  
\end{abstract}


\maketitle

Conical quasiparticle dispersion with valence and conduction bands touching each other at its apex characterizes a new family of materials, Dirac semimetals (DSMs). Such weakly coupled unconventional phase of matter can be realized in both two and three spatial dimensions and graphene stands as a prototype 2D DSM. Although the experimental realization of 3D DSMs remained illusive for long time, very recently their existence has been reported in Cd$_3$As$_2$ \cite{cdas} and Na$_3$Bi \cite{nabi}. In addition, massless Dirac fermionic excitations can be realized at the transition point between 3D strong $Z_2$ topological insulator and trivial band insulator, which, for example, can be tuned by doping or applying external pressure \cite{hassan-cava, ando, hassan-neupane, armitage, TPT-BiT}.

A question of deep and fundamental importance is the stability of DSMs against electron-electron interaction and disorder. Vanishing density of states (DOS) in the vicinity of the band-touching diabolic point leaves DSMs robust against any weak electron-electron interaction in two and three dimensions. However, application of strong external magnetic fields can drive DSMs into fully gapped phases even for arbitrarily weak interactions \cite{roy-kennett-dassarma, roy-sau}. Stability of DSMs against static disorder (e.g. random quenched impurities which are unavoidable in solid state systems) is, however, a subtle and important issue, which we address in this work combining the powerful techniques of renormalization group (RG), scaling analysis, and replica theory. The dimensionality of the \emph{replica averaged} disorder coupling ($\Delta$) is $[\Delta]=2 z -d$, where $z$ is the \emph{dynamical critical exponent} (DCE) and $d$ is the dimensionality of the system. Since, $[\Delta]=-1$ for $z=1$ and $d=3$, any weak short-range disorder is an \emph{irrelevant} perturbation in 3D DSMs. On the other hand, beyond a critical strength (i.e. when the disorder can no longer be considered weak), potential disorder drives DSMs into diffusive metal (DM) \cite{fradkin, goswami-chakravarty, herbut-disorder, shindou-murakami, ominato-koshino, brouwer}, where the quasiparticle lifetime, mean free path as well as the metallic conductivity at $T=0$, are \emph{finite}. However, in 2D DSMs disorder is a \emph{marginally relevant} parameter and the non-interacting system flows toward strong disorder (i.e. even in the presence of infinitesimal disorder) \cite{aleiner-efetov, 2d-Dirac}. Therefore, the disorder-driven DSM-DM QPT is a unique phenomenon in 3D system and we address its quantum critical behavior within the framework of an $\epsilon$-expansion around the \emph{lower critical dimension} $d=2$, where $\epsilon=d-2$. The fact that such a disorder-driven DSM-DM transition exists in three dimensions (but not in 2D) has been known for a long time \cite{fradkin, goswami-chakravarty, herbut-disorder, shindou-murakami, ominato-koshino, brouwer, radzihovsky-gurarie}, but our focus is a deeper understanding of its quantum critical behavior which is still lacking in spite of considerable earlier works. Our main results are: $(a)$ we argue that potential disorder driven QPT is always continuous, $(b)$ critical exponents are obtained to the order $\epsilon^2$, which are most accurate ones in literature, $(c)$ universality class of this QPT is independent of flavor number of Dirac fermions and therefore our study can be germane to disordered Cd$_3$As$_2$, Na$_3$Bi, $(d)$ to two loop order the continuum description of this problem is \emph{renormalizable}, which, however, we expect to hold to all order in perturbative expansion, $(e)$ universality classes of the DSM-DM transitions driven by potential and odd-parity disorders are same, which together constitute a global and broad picture of disorder driven QCP in 3D DSMs.

The nature of DSM-DM transition in 3D has been addressed in various theoretical works. An one loop RG study predicted the DCE $z$ to be $3/2$ and the \emph{correlation length exponent} (CLE) $\nu=1$ near this critical point \cite{goswami-chakravarty}. A numerical analysis also indicated the existence of a sharp transition from DSM to DM at finite disorder strength, albeit with critical exponents inconsistent with the one-loop RG theory within error bars \cite{herbut-disorder}. Existence of DSM-DM transition can also be established from the self-consistent solution of quasiparticle lifetime \cite{shindou-murakami, ominato-koshino} and vanishing ac conductivity \cite{brouwer}. Scaling of vanishing (finite) DOS near the Dirac point in the DSM (DM) phase allows one to extract various exponents near the transition giving $z=1.5 \pm 0.1$, $\nu_{DSM}=0.81 \pm 0.21$ and $\nu_{DM}=0.92 \pm 0.13$ based on a direct numerical simulation \cite{herbut-disorder}. Here $\nu_{DSM}$ and $\nu_{DM}$ are CLEs in DSM and DM phases, respectively. Such large error bars in critical exponents possibly stems from the large uncertainty associated with the location of DSM-DM QCP. The true CLE at the DSM-DM critical point is therefore given by $\nu=\frac{\nu_{DSM}+\nu_{DM}}{2} \approx 0.9$ \cite{herbut-disorder}. A \emph{sigma model} approach to this problem also suggested the existence of such a transition long ago and obtained $\nu^{-1}=d-2$ and $z=2$ \cite{fradkin}. In addition, predictions for finite dc conductivity in weakly disordered DSMs \cite{hosur-vishwanath, radzihovsky-gurarie, ryu-biswas} are in apparent contradistinction with recent numerical calculation \cite{brouwer}. Therefore, even though various works support the existence of a DSM-DM transition at finite disorder strength, exponents reported in Refs. \onlinecite{goswami-chakravarty, herbut-disorder, fradkin} are not in agreement with each other and in this Letter we extend the paeturbative analysis up to two loop order, and unearth various novel and nonperturbative aspects associated with such peculiar QCP. The critical exponents next to the leading order are $\nu^{-1}=\epsilon-\frac{\epsilon^2}{8}$ and $z=1+\frac{\epsilon}{2}-\frac{3}{16}\epsilon^2$.

\begin{figure}[htb]
\includegraphics[width=8.0cm,height=10.0cm]{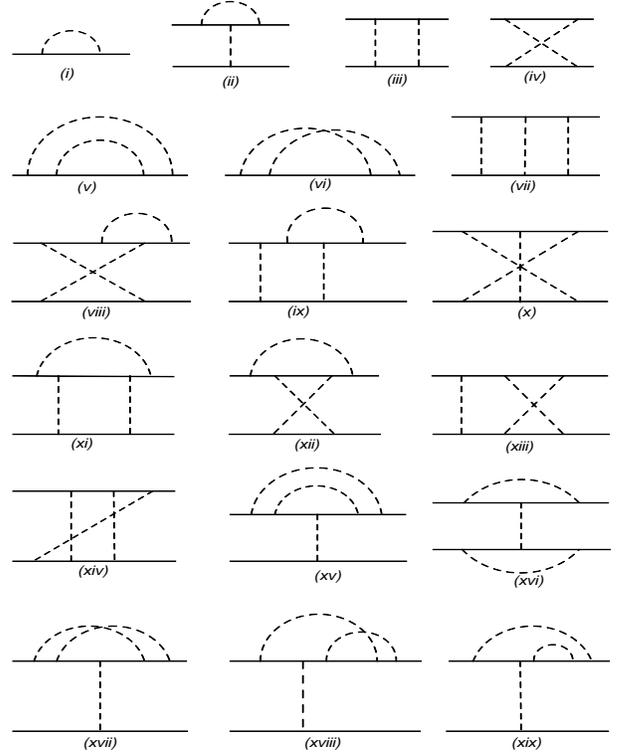}
\caption{One-loop (first row) and two loop diagrams that contribute to the renormalization of the self-energy and disorder vertex in the vanishing replica limit. Solid (dashed) line represents fermions (disorder).}\label{two-loop-disorder}
\end{figure}

The imaginary time ($\tau$) action for noninteracting massless Dirac fermions in $d$-dimensions reads 
\begin{equation}
S_0 = \int d^d\vec{x} d\tau \; \bar{\Psi}(\tau,\vec{x}) \left(\gamma_0 \partial_0 + v \; \gamma_j \partial_j \right) \Psi(\tau,\vec{x}),  
\end{equation}
where a summation over repeated spatial index $j=1, \cdots, d$ is assumed and $\bar{\Psi}=\Psi^\dagger \gamma_0$ as usual. Fermi velocity of the Dirac quasiparticles is $v$. The $\gamma$ matrices satisfy the anticommuting Clifford algebra $\left\{ \gamma_\mu, \gamma_\nu \right\}=2 \delta_{\mu \nu}$ for $\mu,\nu=0,1, \cdots, d$. The fermionic Green's function is $G (k_0, \vec{k})=\left(i \gamma_0 k_0 + i v \gamma_j k_j \right)^{-1}$. Next we add a term $S_D=\int d^d \vec{x} d\tau V(\vec{x}) \left( \bar{\Psi} \gamma_0 \Psi \right)$ that captures the effect of random onsite potential disorder. Performing the disorder averaging we obtain the replicated action 
\begin{eqnarray}\label{replicaS}
\bar{S} = \int d^d\vec{x} d\tau \; \bar{\Psi}_\alpha(\tau,\vec{x}) \left(\gamma_0 \partial_0 + v \gamma_j \partial_j \right) \Psi_\alpha(\tau,\vec{x})  \nonumber \\
- \frac{\Delta}{2} \int d^d\vec{x} d\tau d\tau' \left[ \bar{\Psi}_\alpha \gamma_0 \Psi_\alpha\right]_{(\tau,\vec{x})} \;
 \left[ \bar{\Psi}_\beta \gamma_0 \Psi_\beta\right]_{(\tau',\vec{x})},
\end{eqnarray} 
where $\alpha, \beta$ are replica indices. Disorder averaging is performed assuming a Gaussian white noise disorder distribution, i.e., $\langle \langle V(\vec{x}) V(\vec{x}')\rangle \rangle= \Delta \delta^{d} (\vec{x}-\vec{x}')$. Scale invariance of $\bar{S}$ dictates the dimensionality of fermionic fields $[\bar{\Psi}]=[\Psi]=\frac{d}{2}$, Fermi velocity $[v]=z-1$ and disorder coupling $[\Delta]=2 z-d$. Therefore, in 3D DSMs ($z=1,d=3$) weak disorder is an \emph{irrelevant} perturbation and DSM-DM transition at finite disorder coupling can be accessed using a controlled $\epsilon$-expansion around $d=2$. Relevant diagrams in the vanishing replica limit up to the two loop order are shown in Fig.~\ref{two-loop-disorder}.

We now derive the flow of various parameters in $\bar{S}$ under coarse graining. One loop diagrams in Fig.~\ref{two-loop-disorder} can be evaluated straightforwardly using the Wilsonian shell-RG technique. Nevertheless, we take this opportunity to formulate the $\epsilon$-expansion scheme for this problem and establish the background necessary for two loop calculations. In addition, this exercise allows to contrast the leading order results obtained from $\epsilon$-expansion with other existing studies \cite{goswami-chakravarty, herbut-disorder, fradkin, radzihovsky-gurarie}. Fig.~1$(i)$ gives the leading correction to fermionic self energy 
\begin{equation}
\Sigma\left( i p_0, \vec{p} \right)= \Delta \left(-i p_0 \gamma_0 \right) \Gamma\left( 1-d/2\right)= \frac{2\Delta}{\epsilon} \left(i p_0 \gamma_0 \right), 
\end{equation}
as $d \to 2+\epsilon$ and after rescaling $\frac{p^\epsilon \Delta}{(4 \pi)^{d/2}} \to \Delta$. Here $1/\epsilon$ represents ``$\log$" divergence. On the other hand, the total contribution from diagrams $(iii)$ and $(iv)$ is
\begin{eqnarray}
I_{(iii)} + I_{(iv)} = \frac{4 \cdot \Delta^2}{4 \cdot 2!}  \int \frac{d^d\vec{p}}{(2 \pi)^d} \left[\bar{\Psi}_\alpha \gamma_0 G(0,\vec{p}) \gamma_0 \Psi_\alpha \right] \nonumber \\  
\times \left( \bar{\Psi}_\beta \gamma_0 G(0,\vec{p}) \gamma_0 \Psi_\beta + \bar{\Psi}_\beta \gamma_0 G(0,-\vec{p}) \gamma_0 \Psi_\beta\right) \equiv 0, \quad
\end{eqnarray}
since the relativistic Green's function is \emph{odd} under the reversal of frequency and momenta, i.e., $G(-k_0,-\vec{k})=$ $- G(k_0,\vec{k})$. Therefore, renormalization of the disorder coupling ($\Delta$) arises only from diagram $(ii)$ giving
\begin{eqnarray}
I_{(ii)} =\frac{8 \cdot \Delta^2}{4 \cdot 2!} \int \frac{d^d \vec{q}}{(2 \pi)^d} \frac{q_j (q+p)_n}{q^2 (q+p)^2}=\Delta^2 \Gamma(1-d/2),
\end{eqnarray} 
which yields the renormalization coefficient $Z_\Delta= 1- \Delta \frac{4}{\epsilon}$ as $d \to 2+\epsilon$. $\Sigma(p_0, \vec{p})$ and $Z_\Delta$ lead respectively to the $\beta$-functions (infrared) for $v$ and $\Delta$
\begin{equation} 
\beta_v=v \left(z-1-2 \Delta \right), \: \: \beta_{\Delta}=-\epsilon \Delta + 4 \Delta^2.
\end{equation}
$\beta_{\Delta}=0$ has two solutions: $(a)$ $\Delta=0$ represents stable DSM and $(b)$ DSM-DM transition takes place at critical disorder coupling $\Delta=\Delta_\ast=\epsilon/4$. The CLE is given by 
\begin{equation}
\nu^{-1}= \frac{d \beta_\Delta}{d\Delta} \Big|_{\Delta=\Delta_\ast} =\epsilon.
\end{equation}
Substituting $\Delta=\Delta_\ast$ in $\beta_v$ and keeping $v$ fixed under RG (i.e., $\beta_v=0$), we find the DCE at the transition to be $z=1+ 2 \Delta_\ast=1+ \frac{\epsilon}{2}$. Setting $\epsilon =1$ one obtains $\nu=1$ and $z=3/2$ at the DSM-DM quantum critical point in 3D, which is in agreement with a recent one loop calculation \cite{goswami-chakravarty}. However, it should be noted that $\nu=1$ is an artifact of quadratic approximation in $\beta_\Delta$. Next we show that both $\nu$ and $z$ receive corrections from higher order perturbative expansion.

We now proceed to the next order in perturbation theory and take into account two loop diagrams, shown in Fig.~\ref{two-loop-disorder}. The diagrams $(v)-(vi)$ and $(vii)-(xix)$ renormalize respectively fermionic self-energy ($\Sigma$) and disorder coupling ($\Delta$). Since $G(k_0,\vec{k})$ is an odd function of $k_0$ and $\vec{k}$, contributions coming from $(vii)$ and $(xiv)$, $(viii)$ and $(ix)$, $(x)$ and $(xiii)$, $(xi)$ and $(xii)$ mutually cancel each other. For example, contributions from diagrams $(ix)$ and $(viii)$ together go as
\begin{eqnarray}
I_{(ix)}+ I_{(viii)}=\int_{\vec{k},\vec{q}} \bar{\Psi}_\beta \gamma_0 \left[ G (0,\vec{k}) + G (0,-\vec{k}) \right] \gamma_0 \Psi_\beta \nonumber \\
\times \left[\bar{\Psi}_\alpha \gamma_0 G (0,\vec{k}) \gamma_0 G(0,\vec{k}+\vec{q}) \gamma_0 G (0,\vec{q}) \gamma_0 \Psi_\alpha \right] \equiv 0. \quad
\end{eqnarray} 
Therefore, a general result is that at every order in pertubative expansion each \emph{ladder} diagram is accompanied by an appropriate \emph{crossing} diagram which together give zero net contribution. This cancellation in turn ensures that if the bare replicated model $\bar{S}$ contains only the potential disorder, it does not generate any \emph{new} disorder through loop corrections and $\bar{S}$ remains \emph{closed} under RG to all orders in the perturbation theory. On the other hand, the momentum integral arising from digram $(xix)$
\begin{eqnarray}
I_{(xix)} \propto \int_{\vec{k},\vec{q}} G (0,\vec{k}') \gamma_0 G(0,\vec{q}) \gamma_0 G(0,\vec{k}')
\gamma_0 G(0,\vec{k}') \gamma_0, \nonumber
\end{eqnarray}
where $\vec{k}'=\vec{k}+\vec{p}$, is an odd function of $q$, and thus $I_{(xix)}=0$, which holds even when we keep the external frequency finite. Therefore, renormalization of disorder coupling ($\Delta$) comes from diagrams $(xv)-(xviii)$, which we compute after setting the external frequency to zero.

Diagrams $(xv)$ and $(xvi)$ provide $\log^2$ corrections:  
\begin{eqnarray}
I_{(xvi)} =  \Delta^3 \frac{p^{2 \epsilon}}{(4 \pi)^d} \Gamma\left( 1-d/2\right)^2, 
\end{eqnarray}
and $I_{(xv)}=I_{(xvi)}/2$, whereas $(xvii)$ and $(xviii)$ yield
\begin{equation}
I_{(xvii)}=\Delta^3 \bar{\Gamma}_{j l m n} I_+(p),
I_{(xviii)}=2\Delta^3 \bar{\Gamma}_{j l m n} I_-(p), 
\end{equation}
respectively, where $\bar{\Gamma}_{j l m n}=\gamma_j \gamma_l \gamma_m \gamma_n$. Momentum integrals appearing in $I_{(xviii)}$ and $I_{(xvii)}$ give 
\begin{widetext}
\begin{eqnarray}
I_\pm (p)&=&\pm \bar{\Gamma}_{j l m n} \int_{\vec{k}, \vec{q}} \frac{k_j k_l (k+q \pm p)_m  q_n }{k^2 k^2 (k+q \pm p)^2 q^2}=
\pm \frac{\bar{\Gamma}_{j l m n}}{(4 \pi)^{d/2}} \int_{\vec{k}} \frac{k_j k_l}{(k^2)^2} \left[\frac{\delta_{mn}}{2} \frac{\Gamma\left(1-d/2 \right)}{\left[(k \pm p)^2\right]^{1-d/2}} -\Gamma(2-d/2) \frac{(k \pm p)_n (k \pm p)_m}{\left[(k \pm p)^2\right]^{2-d/2}}\right] \nonumber \\
&=&\pm \frac{p^{2\epsilon}}{(4 \pi)^d}\bar{\Gamma}_{j l m n} \left[\frac{3 \delta_{mn} \delta_{jl}+ \delta_{jm}  \delta_{ln}+ \delta_{jn}  \delta_{lm}}{8 \epsilon} - \frac{ \delta_{mn} \delta_{jl}}{2} \left( \frac{1}{\epsilon^2} +\frac{\gamma_E}{\epsilon}\right)\right]
=\pm \frac{p^{2\epsilon}}{(4 \pi)^d} \cdot \frac{1}{\epsilon} \mp \frac{p^{2\epsilon}}{(4 \pi)^d} \Gamma\left(1-d/2 \right)^2.  
\end{eqnarray}
\end{widetext}
Therefore, the renormalization coefficient of disorder coupling to two loop order is given by
\begin{equation}
Z_{\Delta/2}=1- \Delta \frac{2}{\epsilon} + \frac{\Delta^2}{\epsilon} + \Delta^2 \Gamma\left( 1-d/2\right)^2,
\end{equation}
from which we obtain the $\beta$-function (infrared)
\begin{equation}\label{betadelttwo}
\beta_\Delta=-\epsilon \Delta + 4 \Delta^2-2 \Delta^3.
\end{equation}
The ``$\log^2$" terms in $Z_\Delta$ drop our from the above $\beta$-function. The DSM-DM critical point is now located at $\Delta=\Delta_\ast=\frac{1}{2} \left(2-\sqrt{2} \sqrt{2-\epsilon} \right)$ $=\frac{\epsilon}{4}+\frac{\epsilon^2}{32}+{\cal O}(\epsilon^3)$ near which the CLE
\begin{equation}
\nu^{-1}=\epsilon-\frac{\epsilon^2}{8}+{\cal O}(\epsilon^3) \: \: \Rightarrow \nu= 1.1428 \: \: \mbox{as $\epsilon \to 1$}. 
\end{equation} 
We find yet another solution $\Delta' \gg \Delta_\ast$ of $\beta_\Delta=0$ (besides the trivial one at $\Delta=0$). However, the appearance of such \emph{putative} fixed point is only an artifact of solving a cubic equation $\beta_\Delta=0$. Stability of DSM-DM QCP against two-loop corrections along with the fact that potential disorder does not generate any new disorder vertex at any order in perturbation theory ensures that such QPT is always continuous or second-order in nature.

In order to find the correction to the DCE to the same order one needs the self-energy renormalization coming from diagrams $(v)$ and $(vi)$ in Fig.~\ref{two-loop-disorder}. Diagram $(v)$ gives 
\begin{eqnarray}
I_{(v)}= (-i p_0 \gamma_0) \Delta^2 \frac{p^{2\epsilon}_0}{(4 \pi)^d} \bigg[ \frac{1}{2} \Gamma\left(1-d/2\right)^2 + \frac{1}{\epsilon}\bigg].
\end{eqnarray}
Computation of $(vi)$ is most difficult task which reads as
\begin{eqnarray}
I_{(vi)}=\Delta^2  \int_{\vec{k},\vec{q}} \gamma_0 G(p_0, \vec{p}+\vec{q}) \gamma_0 G(p_0, \vec{k}+\vec{q}) \gamma_0 G(p_0, \vec{k}) \gamma_0 \nonumber
\end{eqnarray}
\begin{eqnarray}
=\frac{\Delta^2}{(4 \pi)^d} \left[-\frac{1}{\epsilon} - \frac{1}{\epsilon}\left(-1+ \gamma_E \right)\right] \int^1_0 dx \left[x(1-x)\right]^{1-d/2} \times \quad \nonumber \\
 \int^1_0 dy  \frac{ y^{-d/2} \left(i k_0 \gamma_0 -y i k_j \gamma_j \right)}{\left[ x y (1-x)(1-y)p^2 + \left(y+ x (1-x) (1-y) \right) p^2_0 \right]^{2-d}} \nonumber
\end{eqnarray} 
\begin{eqnarray}
= \Delta^2 \left(-i p_0 \gamma_0 \right) \left[\frac{1}{2} \Gamma\left(1-\frac{d}{2} \right)^2 +\frac{4}{\epsilon}\right] + \frac{\Delta^2}{\epsilon}  \left(-i p_j \gamma_j \right), \quad
\end{eqnarray}
after taking $\Delta p^\epsilon/(4 \pi)^{d/2} \to \Delta$. $I_{(vi)}$ can be computed analytically for zero external frequency ($p_0=0$) and momentum ($\vec{p}=0$) separately that respectively give the spatial and temporal parts of the self energy. When $p_0, p_j \neq 0$, the final integration over the Feynman parameters $(x,y)$ has to be carried out numerically with the on-shell condition $p_0=|\vec{p}|$ for the external fermions. These two procedures, however, lead to identical self energy corrections. Hence, the total self-energy is
\begin{eqnarray}
\Sigma(i p_0, \vec{p})&=&(-i p_0 \gamma_0) \bigg[ \Delta \Gamma\left(1-\frac{d}{2}\right) + \Delta^2 \Gamma\left(1-\frac{d}{2}\right)^2 \nonumber \\
&+&  \Delta^2 \left( \frac{4+1}{\epsilon} \right) \bigg] + (-i p_j \gamma_j) v  \Delta^2 \; \left( \frac{1}{\epsilon} \right),
\end{eqnarray}
which leads to the $\beta$-function for Fermi velocity $(v)$
\begin{equation}\label{betavtow}
\beta_v=v \left(z-1 - 2 \Delta+ 4 \Delta^2 \right).
\end{equation}
 Terms proportional to ``$\log^2$" in $\Sigma(ip_0,\vec{p})$ disappear from $\beta_v$. Cancellation of ``$\log^2$" terms in the $\beta$-functions in Eqs.~\ref{betadelttwo} and \ref{betavtow} ensures that the theory in Eq.~\ref{replicaS} is renormalizable at least to two loop order. However, we believe that theory remains renormalizable to all order in perturbation theory. Substituting $\Delta=\Delta_\ast$ in $\beta_v$ and setting $\beta_v=0$, we obtain the DCE ($z$) at DSM-DM quantum critical point to be 
\begin{equation}
z=1+\frac{\epsilon}{2}-\frac{3}{16} \epsilon^2 + {\cal O}(\epsilon^3) \: \: \Rightarrow z = 1.3125 \quad \mbox{as $\epsilon \to 1$}.
\end{equation}

Therefore, both CLE and DCE receive significant corrections from higher loops and are noticeably different from their values predicted from theories based on one loop RG \cite{goswami-chakravarty}, sigma model \cite{fradkin} and numerical simulation \cite{herbut-disorder}. The smallness of the coefficient of $\epsilon^2$ in $\Delta_\ast$, $\nu$ and $z$ justifies the inclusion of corrections to these quantities coming from two loop diagrams. It is worth pointing out that critical exponents at DSM-DM critical point is insensitive to the flavor number of Dirac fermions ($N$). Any diagram that contains fermionic loop, which gives contribution proportional to $N$, vanishes in vanishing replica limit. Therefore, our study is pertinent for DSMs with arbitrary number of Dirac cones as well as disordered Cd$_3$As$_2$, Na$_3$Bi. However, it should be noted that $\beta$-functions in Eqs. (\ref{betadelttwo}), (\ref{betavtow}) manifest oscillatory dependence in powers of $\Delta$. Therefore, actual critical exponents near DSM-DM quantum critical point can only be pinned after performing an infinite order resummation of such series, which may require the notion of few additional higher order terms in $\beta_\Delta$ and $\beta_v$. Nevertheless, exact numerical solutions and experiments can provide valuable insights into the nature of such transition. Our work clearly establishes this problem as an interesting and open quantum critical problem of fundamental importance, and hopefully will stimulate numerical as well as experimental works in this direction in near future.

 DSMs can also suffer from the presence of various other time-reversal symmetric, such as random mass, spin-orbit and odd-parity disorders. DSMs in the presence of mass disorder of arbitrary strength remains stable, but display a QCP at finite odd-parity disorder coupling \cite{goswami-chakravarty}. The odd-parity disorder coupling is captured by a terms $S_{OP}=\int d^3\vec{x} d\tau V(\vec{x}) \bar{\Psi} \gamma_0 \gamma_5 \Psi$, where $\{\gamma_5,\gamma_\mu \}=0$. Otherwise, contributions from all diagrams shown in Fig.~\ref{two-loop-disorder} remain unchanged after taking $\gamma_0 \to \gamma_0 \gamma_5$. Therefore, quantum critical behavior near DSM-DM QCP driven by potential and odd-parity disorders are characterized by identical set of exponents $(\nu, z)$.

The exponents $\nu$ and $z$ govern the scaling behavior of various physical quantities near the critical point \cite{HJR}. For temperatures much smaller than the bandwidth, specific heat assumes the scaling form  
\begin{equation}
C_v=T^{d/z} v^{-3} H\left( \frac{T}{\delta^{\nu z}} \right),
\end{equation}
where the parameter $\delta= (\Delta_\ast-\Delta)/\Delta_\ast$ measures the deviation from the critical point ($\Delta_\ast$). For small arguments the universal scaling function $H(x) \sim x^{d(z-1)/z}$, so that we recover $T^3$ dependence of specific heat in 3D DSMs. Identifying the proportionality constant in $C_v$ as $[v(\delta)]^{-3}$ we also obtain the scaling function for the Fermi velocity $v(\delta)=v \; \delta^{\nu (z-1)}$. In the DM phase $H(x)\sim x^{1-d/z}$ which in turn gives $T$-linear specific heat. In the quantum critical regime $H(x)$ is an universal function, and $C_v \sim T^{d/z}$. Similarly, the universal scaling function for DOS $(\rho)$ is given by \cite{herbut-disorder}
\begin{equation}
\rho(E)=\delta^{(d-z)\nu} F\left(\frac{|E|}{\delta^{\nu z}} \right),
\end{equation}
when the energy $E$ is much smaller than bandwidth. On the DSM side the universal function $F(x) \sim x^{d-1}$ for small argument, yielding $\rho(E)\sim E^{d-1}$. In the DM phase, on the other hand, $F(x) \sim x^0$, giving a finite DOS at the Dirac point. Therefore, the DOS may serve the purpose of an order-parameter across the DSM-DM transition \cite{herbut-disorder, nandkishore}. Residue of the quasiparticle pole remains finite in the entire DSM phase, which, however, vanishes smoothly as $\delta \to 0$ \cite{HJR} and may as well serve as an order-parameter in the DSM side of the transition.

To summarize, we address the quantum critical behavior of onsite potential disorder driven DSM-DM transition in three dimensions within the framework of an $\epsilon$-expansion around the lower critical dimension $d=2$. We show that critical exponents $\nu$ and $z$ receive significant corrections from higher loops and together they determine the scaling of various measurable quantities such as specific heat $(C_v)$ and DOS ($\rho$). However, we have neglected the effects \cite{nandkishore, arovas, das-sarma-graphene, skinner, halperin} of puddles, density inhomogeneities, Lifshitz tail and Griffiths physics in this work focusing instead on the quantum critical aspects of the DSM-DM QPT which are directly accessible to RG analyses. We strongly believe that these additional subtle effects do not fundamentally affect our quantum critical theory.

This work is supported by NSF-JQI-PFC and LPS-CMTC. B. R. thanks I. F. Herbut, V. Juri\v ci\' c and P. Goswami for useful discussions. Authors acknowledge discussion with Kostya Kechedzhi at early stages of this work. B. R. is thankful to Max-Planck Institute for Complex Systems, Dresden, Germany, for hospitality during the workshop ``Topology and Entanglement in Correlated Quantum Systems" where part of the manuscript was finalized.

\end{document}